\documentclass{article}

\usepackage{arxiv}

\usepackage{lineno}
\modulolinenumbers[5]

\usepackage{graphicx,booktabs} 
\usepackage{url,lineno,microtype,subcaption,hyperref}
\usepackage{float}
\usepackage{verbatim} 
\usepackage{multirow}
\usepackage{booktabs}
\usepackage{textcomp}
\usepackage{siunitx}
\usepackage{amsmath}
 \usepackage[applemac]{inputenc} 
 \usepackage{enumitem}
 \usepackage{adjustbox}
\usepackage{xcolor}

\usepackage{amsfonts}       
\usepackage{nicefrac}       
\usepackage{microtype}      

\hypersetup{
    colorlinks,
    citecolor=green,
    anchorcolor=green,
    filecolor=blue,
    linkcolor=blue,
    urlcolor=blue
}
\newcommand{\ts}{\textsuperscript}

\title{Evaluation of a 
Vision-to-Audition Substitution System 
 that Provides 2D WHERE Information and Fast User Learning}

\author{Louis Commère, Jean Rouat\\
Université de Sherbrooke\\
2500 Boulevard de l'Université, Sherbrooke, QC J1K 2R1, Canada\\
\texttt{ \{louis.commere,jean.rouat\}@usherbrooke.ca }
\And
Sean U.N. Wood\\
InteraXon, Inc. \\
 555 Richmond St W suite 900, Toronto, ON M5V 3B1, Canada\\
\texttt{jean.rouat@usherbrooke.ca}
}

\begin{document}
\maketitle

%





\begin{abstract}
Vision to audition substitution devices are designed to convey visual information through auditory input. The acceptance of such systems depends heavily on their ease of use, training time, 
reliability and on the amount of coverage of online auditory perception of current auditory scenes. Existing devices typically 
require extensive training time or complex and computationally demanding technology. The purpose of this work is to investigate the learning curve for 
a vision to audition 
substitution system that provides simple location features. 
Forty-two blindfolded users participated in 
experiments 
involving location and navigation tasks. Participants had no prior experience with the system. For the location task, participants had to locate 3 objects on a table after a short familiarisation period (10 minutes). Then once they understood the manipulation of the device, they proceeded to the navigation task: participants had to walk through a 
large 
corridor 
without colliding 
with obstacles randomly placed on the floor. Participants were asked to repeat the task 5 times. 
In the end of the experiment, each participant had to fill out a questionnaire to 
provide feedback. 
They were able to perform 
localisation and navigation effectively after a short training time 
with an average of 10 minutes. Their navigation skills greatly improved across the trials.
\end{abstract}

\keywords{sensory substitution, vision to audition, obstacles avoidance, navigation, embedded systems, sonification}



\section{The Elsevier article class}

Sensory 
substitution is a mechanism 
in which one sense can stimulate and consolidate the brain activity of another sense, given the extremely plastic nature of the brain~\cite{bach1987brain,kokovayNeuron2008}. 
Initial research conducted by
 ~\cite{bachYRita1969} established that the 
back can be used to mediate visual stimulus to the brain. Trained blind people can 
navigate and ``visually perceive'' 
features of the environment via electric pulses on the tongue that encode camera images~\cite{bachYRitaTrends2003,ptitoBrain2005}.
The substitution from vision to audition has also been studied and Sensory Substitution Devices (SSDs) have been designed. 
We are interested in the potential of such systems to provide information on the location of the elements of a visual scene, referred to in this paper as ``where'' information. 
\emph{The vOICe}~\cite{Meijer1992} is the first 
known vision to audition system 
originally developed by P. Meijer in 1992.
It 
turns greyscale images into sound by scanning them 
from left to right. 
It uses a mapping of the positions and grey values of pixels to frequency and amplitude short-time sine tones. 
The potential of~\emph{the vOICe}  to provide ``where'' information of the visual scene was investigated.
~\cite{auvray2007learning} evaluated the performance of blindfolded sighted participants.
Within approximately 5 hours, participants could locate a black target on a white table.~
~\cite{proulx2008seeing} observed that people who continuously used the~\emph{vOICe} for 21 days in an everyday life natural environment had much better results for a location task than people  trained 4 hours a day in a laboratory.
~\cite{Brown2011} experimented with~\emph{the vOICe} under various experimental conditions. Within 3 hours, blindfolded sighted participants were able to locate objects placed in one of 9 pre-determined positions and to slowly navigate a route around four obstacles.  

Other major vision to audition substitution systems were later designed. The Prosthesis for Substitution of Vision by Audition (\emph{PSVA}) developed by Capelle
~\cite{Capelle1998} holistically encodes the position of pixels with frequencies of sine tones with a greater  resolution in the centre to mimic the fovea. 
~\cite{Renier2010}  showed that early blind and sighted blindfolded participants mastering the \emph{PSVA} (15 hours prior training) could locate - with an additional training of approximately 2 hours - white cubes placed in one of the 9 pre-determined 
locations 
on a table. 
\emph{The Vibe} developed by
~\cite{Auvray2006}, encodes positions and values of virtual receptive fields 
 with interaural disparity, frequency, and amplitude of sine tones.  
The potential of \emph{The Vibe} to provide ``where'' information was evaluated in a navigation experiment conducted over 4 days by
~\cite{durette:inria-00325414} and in a location task conducted by
~\cite{hanneton:hal-00577751} (training time not specified). 

A more recent category of vision to audition devices involves the use of depth sensors, and shows potential in terms of required training time and performance for navigation or localisation
~\cite{stoll2015navigating,aladren2016navigation,gholamalizadeh2017sonification,pourghaemi2018real,Skulimowski2018}. In the work of
~\cite{stoll2015navigating}, blindfolded sighted participants were able to navigate in unknown path after only 8 minutes of training.
~\cite{pourghaemi2018real} showed that 45 minutes of training allowed blindfolded sighted participants to locate and differentiate 6 objects. 
However, such systems currently require external 3D sensors and are computationally demanding which make them expensive and difficult to use outside of the laboratory setting.
~\cite{maidenbaum2014} observe that most SSDs are  mainly 
  designed for use in research laboratories 
  and are not effective for visual rehabilitation. 
We 
argue that the acceptance of visual to sound substitution systems depends heavily on the ease of use and on the required training time. This 
  is supported by
  ~\cite{Dakopoulos2010} who suggest that an efficient SSD should be \textit{``easy to use (operation and interface not loaded with unnecessary features) and without the need of an extensive training period''} and by
  ~\cite{BERTRAM2016234}  who \textit{``believe that improving the efficiency of learning to use an sensory substitution and augmentation device would reduce the effort, time, and money invested by patients and researchers, and potentially makes their use as assistive technology more appealing''}.
The required training time (depending on the task) for simple devices\footnote{Devices that do not rely on 3D representations of the visual scene} to provide ``where'' information is comprised (depending on the experimental procedure) between 3 hours in the best case but can take up to 3 weeks. 
The issues with current systems that are based on depth sensors  
are their complexity, power requirement, and high price. 
 
 In this work, we evaluate the potential of a low-cost sensory substitution application for iPods, iPhones and iPads, named \emph{See Differently}~\cite{rouatICAD2014}.
  The system is based on an easy to interpret sonification that provides the ``where'' information of the visual scene 
   and can be used for long periods of time without the need to be connected to the cloud. Users can interpret the scene and determine their strategy to increase their understanding of the scene.  

\emph{See Differently} 
encodes salient parts of the visual scene before the transformation into sounds. 
  A machine learning analysis of visual scenes could 
  have been implemented to provide sophisticated cues to the user (and eventually use a synthetic voice to describe the environment). 
But such an approach is usually not general enough to be used in a novel environment and we prefer to use a simple device and let participants interpret the auditory information.

The SSD was designed for late blind people who maintain their spatial abilities and spatial cognitive map from their early visual experiences~\cite{dormal2012}. Since they adopt similar strategies to those used by sighted people when performing spatial tasks~\cite{ungar200013,Pasqualotto2013},
we evaluated the learning curve of \emph{See Differently} 
on 46 blindfolded sighted participants (42 individuals, 4 participated 2 times). 
We also conducted an experiment with one congenitally blind participant to have an insight into the potential of the system for such a population. 
 The learning curve refers to how fast the sensory-motor system will efficiently learn a new internal model~\cite{wolpert2011principles} defining new mappings between the movement of the hand holding the device exploring the 3D space, and the new auditory stimulation provided by the SSD. 

We 
present an experimental protocol 
involving 2 tasks. During the first task, participants became familiar with the system and had to locate (by pointing with the index) 3 
objects on a table. During the second task, participants had to navigate through a corridor while avoiding obstacles randomly placed on the ground. 
The performance of the location task is evaluated by measuring the pointing error distance and the time taken to find the 3 objects. The evolution of the performance of the navigation task is evaluated with the number of objects that participants did not detect
on their path and the time they took to complete the courses. 
All participants were able to perform the location and navigation tasks effectively after a short training time (approximately 10 minutes). Performance of the participants for the navigation task significantly improved during the experiments. Interestingly, different strategies were used by the participants which shows great flexibility allowed by the chosen sonification method. Results indicate that a flexible SSD with low computational cost and with short training time requirement is feasible by not overloading the audition and by carefully choosing the transmitted information. 

\section{See differently description}

\begin{figure}[h!] 
   \centering
   \includegraphics[width=0.5\textwidth]{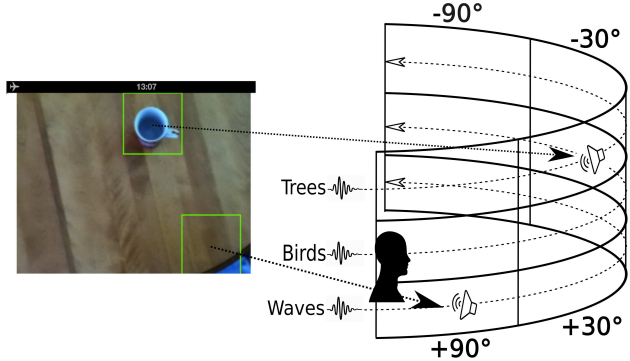} 
   \caption{Principle of See Differently. 
  Camera images are divided into 12 cells of equal size. Each cell is encoded with a different spatialized sound depending on its position. The vertical position of the cell is encoded by the type of the sound (birds, tree or waves sound) and the horizontal position of the square is encoded with the azimuth of the sound. Sounds are played when the image processing algorithm detects enough salient pixels in each cell.
    }
   \label{fig:voirDifferemment}
\end{figure}

Figure~\ref{fig:voirDifferemment} illustrates how the proposed SSD sonifies the 2D video stream of the device's camera. Since the system uses 2D images as input data, distance information is not provided directly to the user. However, the user can understand the 3D environment by actively exploring the space with the hand-held device. 
The image of the camera is divided into 12 cells that are respectively associated with specific prerecorded sounds. Each sound (12 different sounds) is  spatialized via convolution with Head Related Transfer Function (HRTF) filters associated to the center of each square and having respectively azimuthal ($A_z$) and elevation values ($E_l$). Due to the poor elevation encoding with non-individual tunings of HRTF~\cite{stitt2019auditory}, different types of sounds are also used to encode the vertical position of the cells. If the image processing algorithm detects a number of salient pixels in a cell above a certain threshold, the sound associated with this given cell is triggered and looped. When the number of detected salient pixels is below the threshold, the sound stops. The user hears the combination of all sounds triggered by their corresponding cells. The averaged update time of the complete process (image acquisition, neural network processing, and sound playing or stoping) is between 30ms and 45ms (iPod touch, 5\ts{th} generation) and depends on the CPU load of the device. It is dominated by the processing time of the neural network. On more recent devices, the update could be faster but is limited to the acquisition time of the camera.  
 A pair of headphones is used for binaural hearing.
 Details of the image processing algorithm and the sound generation strategy are given in Section~\ref{subsec:imageProcessing},~\ref{sec:areaOfInterest} and~\ref{sec:soundgeneration}.


\subsection{The Image Processing Algorithm}\label{subsec:imageProcessing}
The~\emph{See Differently} SSD identifies salient features in the images of the device's camera with 
 a filtering implementation of a network of neurons~\cite{lescalISSPA2012} 
 and plays sounds depending on the position of the most salient areas\footnote{A video demonstration of the device with a configuration of notes playing is available under
 \href{https://www.gel.usherbrooke.ca/rouat/publications/videoSeeDifferently2019_10.mov}{this link}. 
 }. 
 As the neural network implements a non-linear filter to enhance salient features, no training is required. These features are assumed to be areas with strong
gradients, contrasts, and textures~\cite{lescalISSPA2012}.

To run 
 efficiently on low power processors (iPod touch, 5\ts{th} generation) without the requirement of external resources, only one iteration of the original algorithm 
 is used instead of the recommendation of 4 iterations.
 Weights are positive and encoded in a table lookup on 8 bits. States of neurons are limited to the interval $[0,1]$ so that the output of the network can be used as a mask applied to the original images\footnote{Strong contrast and large local gradients areas are mostly
 enhanced by the neural network, with the mask's values equals to 0.}. 

\begin{figure}[h!]
   \centering
   \includegraphics[width=0.5\textwidth]
   {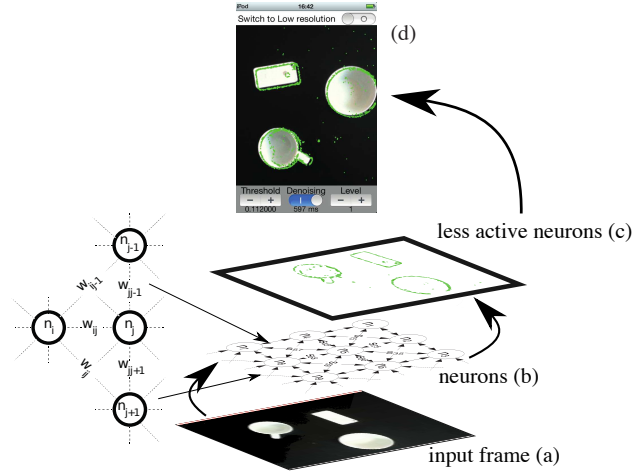}
   \caption{Greyscale input frames are presented as input (a) to the network of neurons. The neural network (b) simultaneously reveals contours, contrasts, and textures. They are encoded into neurons with the smallest state (illustrated in (c) as green). The same output is also plotted and superimposed on the screen of the handheld for comparison purposes (d). The small shift between the input image and the output result is due to the movement of the handheld while capturing the screenshot.}
   \label{fig:analyseImageParHandHeld}
\end{figure}

Each pixel from the camera images is quantized into 256 greyscale levels and encoded into the weights of the neural network (Fig.~\ref{fig:analyseImageParHandHeld}). 
Parameters and a description of the original algorithm of~\cite{lescalISSPA2012} are given in appendix~\ref{appendix:neuralNet}.

\subsection{Determination of areas of interest}\label{sec:areaOfInterest}
 \begin{figure}[h!]
   \centering
   \includegraphics[width=0.4\textwidth]
   {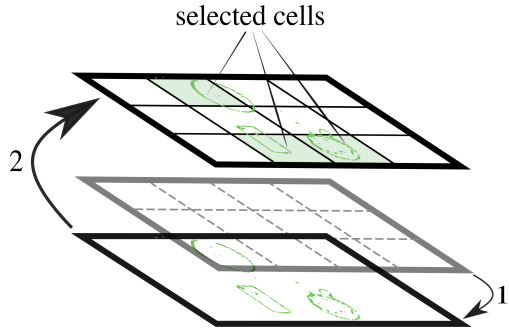}
   \caption{A 3 x 4 grid is superimposed on the neural network (step 1). The cells with the greatest number of non-active neurons are then selected (step 2).}
   \label{fig:positionnementDeLaGrille}
\end{figure}

\begin{figure}
\centering
\begin{minipage}[h]{0.3\textwidth}
\begin{subfigure}[h]{\linewidth}
    \centering  
    \includegraphics[width=0.7\linewidth,height=3cm]{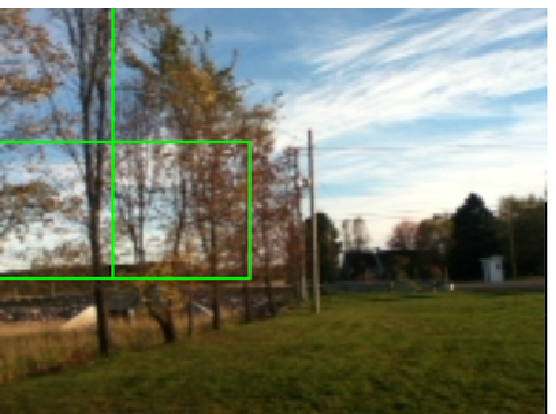}
    \centerline{(a) Trees}
\end{subfigure}
\begin{subfigure}[h]{\linewidth}
\vspace{2mm} 
  \centering
\includegraphics[width=0.7\linewidth,height=3cm]{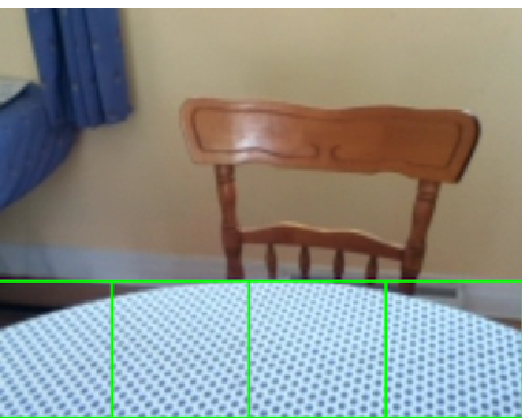}
\centerline{(c) Tablecloth}
\end{subfigure}
\end{minipage}
\begin{minipage}[h]{0.3\textwidth}
\begin{subfigure}[h]{\linewidth}
  \centering
  \includegraphics[width=0.7\linewidth,height=3cm]{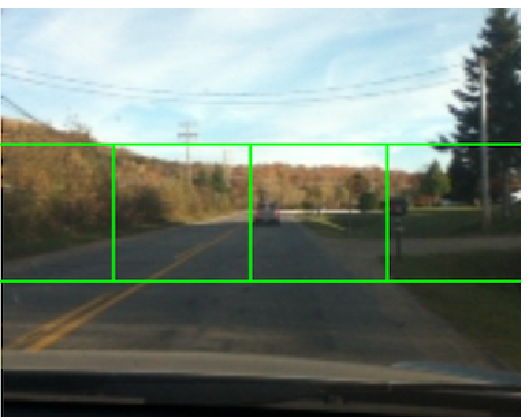}
   \centerline{(b) On the road} 
\end{subfigure}
\begin{subfigure}[b]{\linewidth}
\vspace{2mm} 
  \centering
\includegraphics[width=0.7\linewidth,height=3cm]{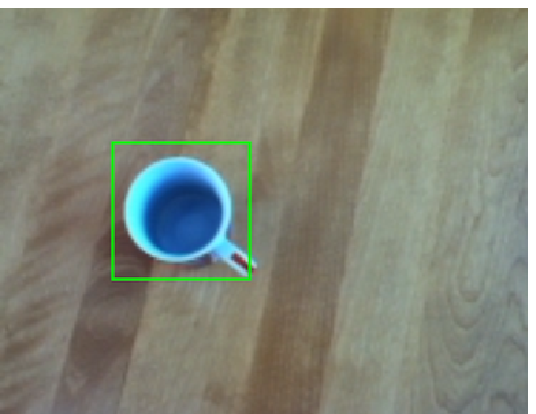}
  \centerline{(d) Cup}
\end{subfigure}
\end{minipage}
\caption{Examples of areas of interest found by the handheld device.}
\label{fig:results}
\end{figure}

The number of salient points is computed for each cell in a 3x4 grid superimposed on the image processing output (Fig.~\ref{fig:positionnementDeLaGrille}). If the number of salient points for a given cell exceeds a predetermined threshold, the cell from the grid is declared to be associated with an area of interest in the original image. The threshold is automatically determined by the size of the handheld screen and the number of cells. Examples of areas of interest are shown in Figure~\ref{fig:results}.
%

\subsection{Sound generation}\label{sec:soundgeneration}
The 12 cells from the grid were mapped to 12 virtual sound sources located according to their respective azimuths $A_z$ and elevations $E_z$  (Fig.~\ref{fig:voirDifferemment}). Four azimuths (+90, 30, -30 and -90 degrees) and three elevations (-40, 0 and 45 degrees) angles were mapped respectively to each virtual source. 
   
      The 2D illusion is obtained by convolving the monophonic sound with the stereo HRTF filters corresponding to the location ($A_z$,$E_l$). The HRTF filters were defined for the 4 azimuths and 3 elevations associated with the 12 cells of the grid. The filters are taken from the KEMAR manikin of the CIPIC database~\cite{CIPIC2001}. 
Three environmental noises (waves on the beach, bird songs, and tree cracking) are played in function of the elevation, thus encoding elevation not only with the HRTF but also within the type of sounds. The 4 azimuth angles of +90, 30, -30 and -90 degrees were used to encode the virtual azimuths of each 3 sounds in the headphones. 
The original 3 environmental sounds have approximately the same loudness (estimated from listening experiments from the authors).

\section{Protocols and Experiments}

\begin{figure}[H]
\centering
\begin{minipage}{.25\textwidth}
\begin{subfigure}{0.99\linewidth}
    \centering  \includegraphics[width=0.9\linewidth,height=3.2cm]{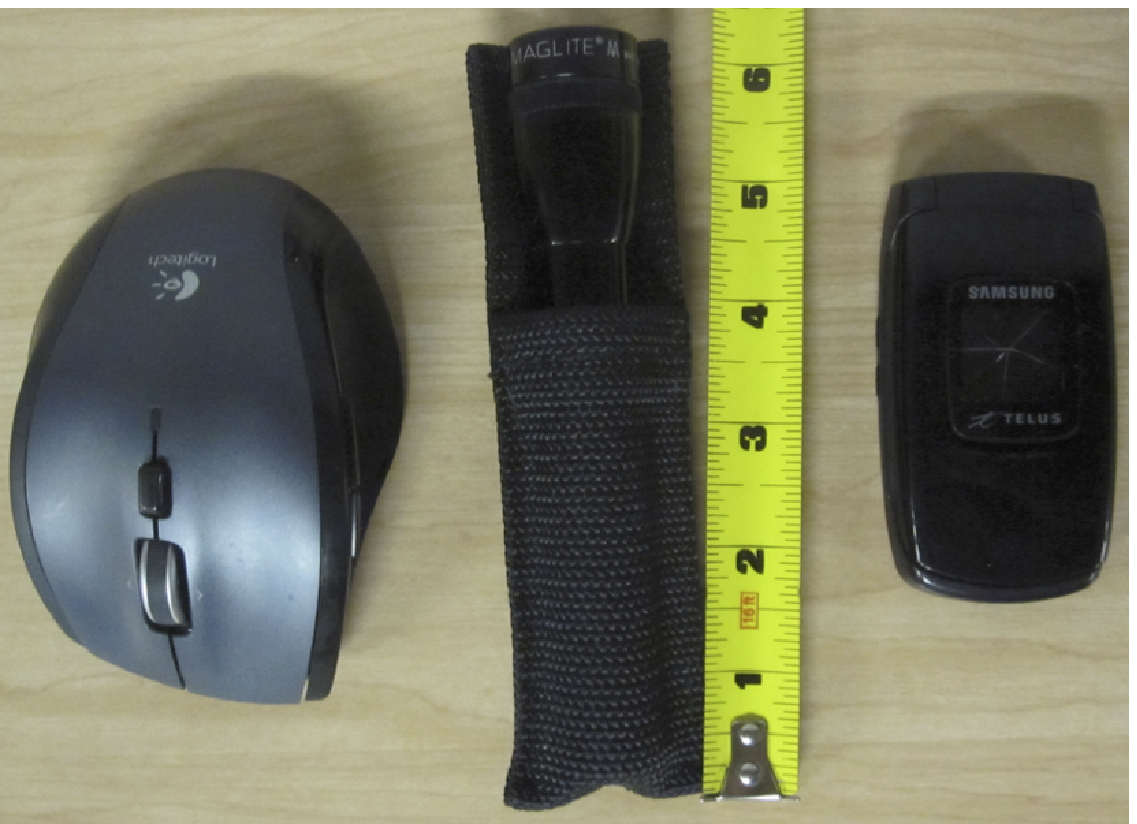}
    \centerline{(a)}
\end{subfigure}
\end{minipage}
\begin{minipage}{.25\textwidth}
\begin{subfigure}{0.99\linewidth}
    \centering  \includegraphics[width=0.9\linewidth,height=3.2cm]{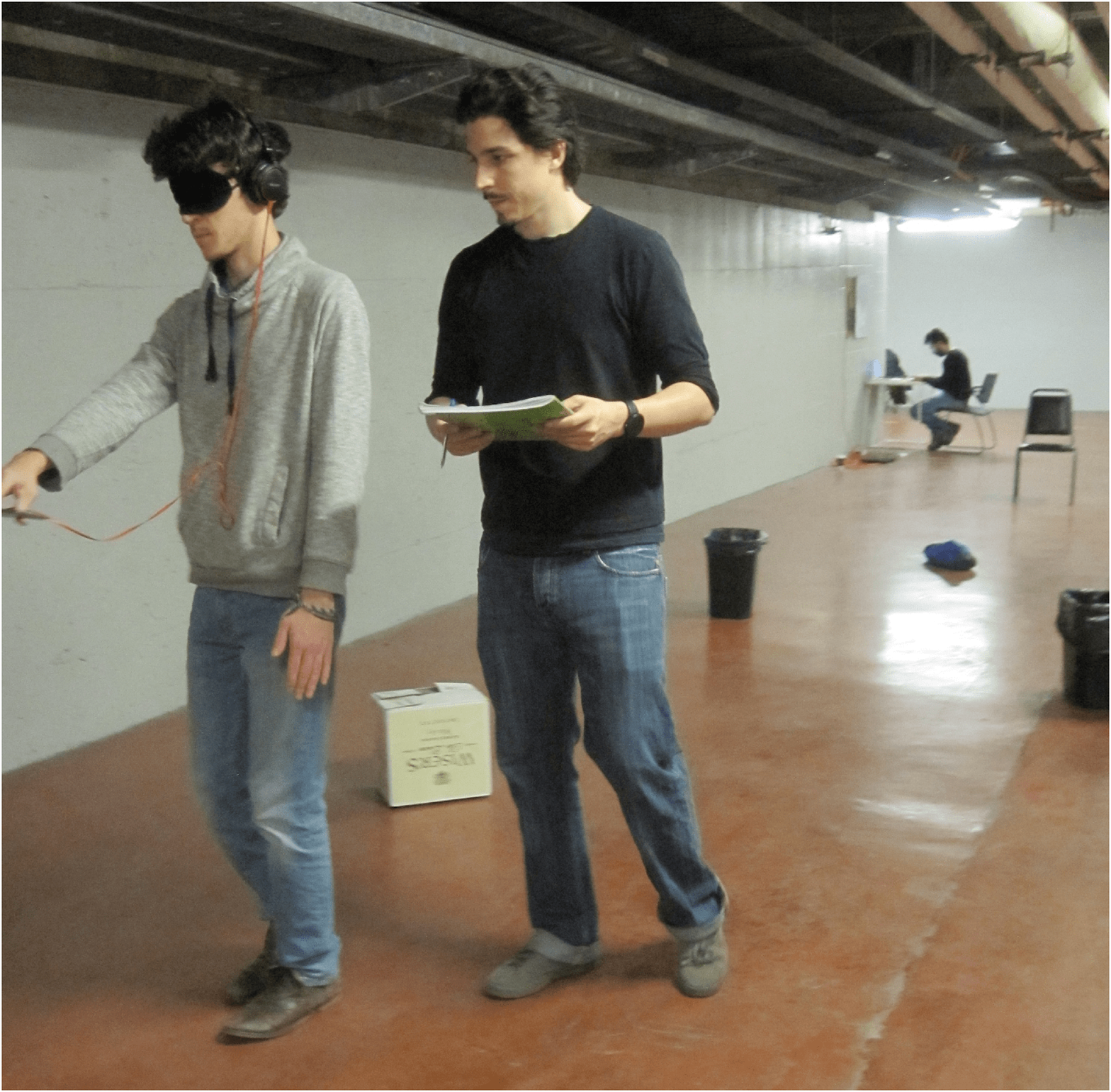}
    \centerline{(b)}
\end{subfigure}
\end{minipage}
\begin{minipage}{.25\textwidth}
\begin{subfigure}{0.99\textwidth}
    \centering  \includegraphics[width=0.9\linewidth,height=3.2cm]{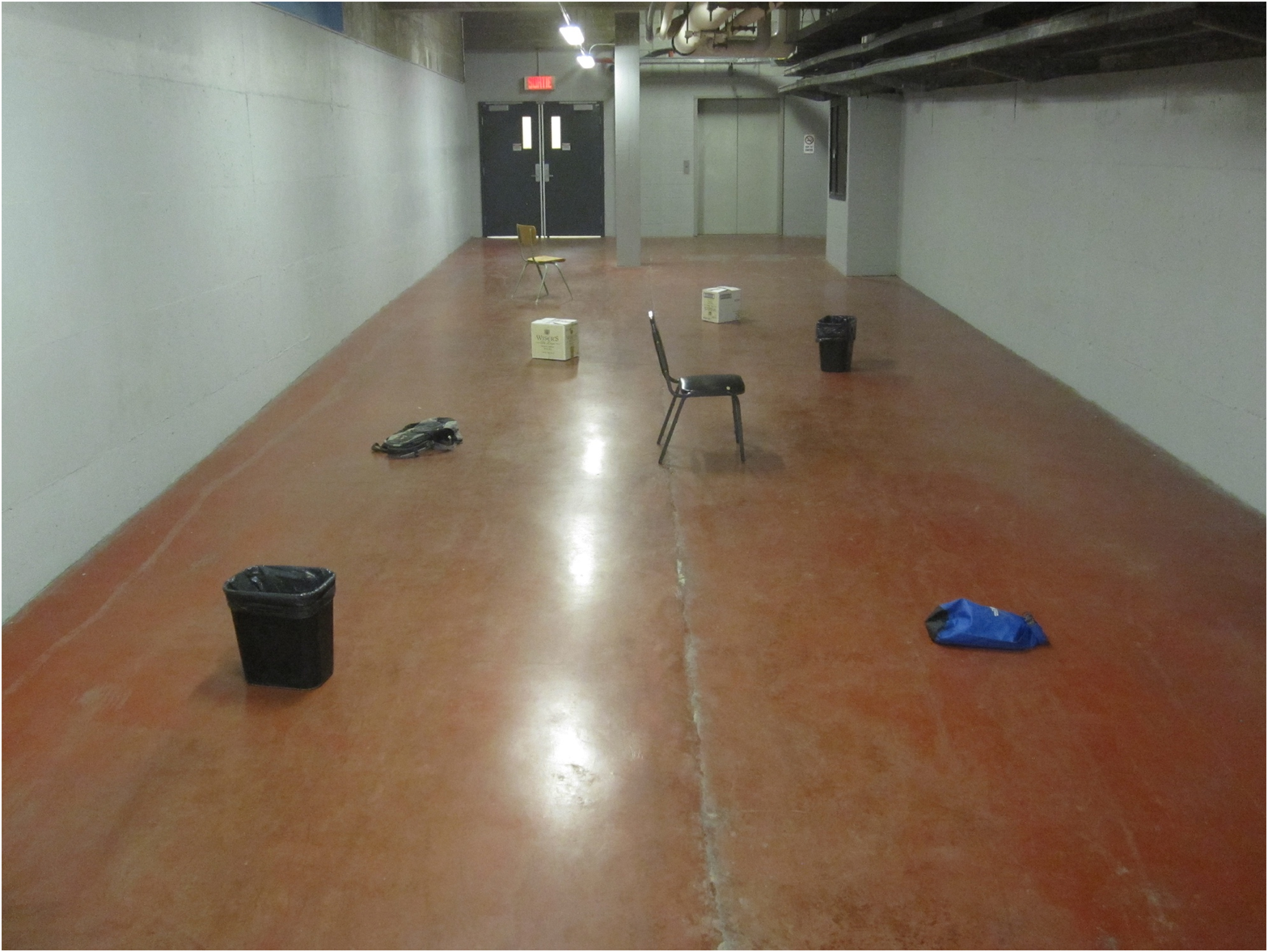}
\centerline{(c)}
\end{subfigure}
\end{minipage}
  \caption{familiarisation (a), navigation (b), and corridor (c). (a): A computer mouse, flashlight, and cellular phone have to be found on a table. (b) The participant scans the environment by moving the handheld around, while the assistant takes notes of the strategy being used by the participant. (c) : The corridor and the 8 obstacles for the navigation task.}
   \label{fig:experiences}
\end{figure}

Experiments are conducted around 2 tasks. 
The first involved familiarisation and localizing objects with the device and the second involved navigation. Experiments were conducted on three separate days spread over one year.
We report results with 46 blindfolded sighted participants (42 individuals, 4 participated 2 times). 
%
 One experimental session, including the 2 tasks, lasts between 30 minutes and 1 hour depending on the time taken by the participants to complete the tasks.
 Participants could either use their own headphones or the Sony brand pair we provided. We designed the system to be, if possible, independent on the headphones. In fact, users should fill comfortable with the use of their own headset.
 They did not have any distinctive health-related problems.
For the first task, participants had to locate 3 objects on a table (Fig.~\ref{fig:experiences}a). Once they understood the manipulation of the device, they could proceed to the navigation task (Fig.~\ref{fig:experiences}b,c). Both tasks were approved by the ethical committee from Letters and Human Sciences from Universit\'{e} de Sherbrooke under reference number 2014-85/Rouat. Participants had to answer a questionnaire after completing the 2 tasks. Based on the ethical certificate, each participant received the same amount of financial compensation to cover their expenses and participation. None of them were relatives or friends of the authors.


\subsection{familiarisation and localisation}
\subsubsection{Protocol} 
\label{sec:FamProtocol}
Participants were seated in front of a table, blindfolded, and asked to gradually increase the sound level (set at the minimum before the experiment) to a comfortable level during familiarisation with the device. They were then asked to locate three objects: a wireless computer mouse, a mobile phone, and a flashlight. They were allowed to explore and touch the table and the objects while 
listening to the sounds played by the device through headphones. They did not have to identify the objects. After the short exploration phase (lasting on the average 10 minutes), we measured their accuracy and speed in pointing at the 3 objects placed simultaneously on the table. We proceeded as follows:
\begin{enumerate}[label=(\roman*)]
\item The 3 objects were placed randomly on the table within a range of 5 centimeters to 1 meter from the participant.
\item The participant was asked to point to the first object they find;
\item The participant was then asked to advance their index finger until making contact with the object, or the table if the object was missed;
\item If the object was missed, the distance between the center of the object and the index finger of the participant was measured (called the pointing error distance);
\item Then the participant proceeded to find the other objects following the same procedure from
 (iii) and (iv) until the 3 objects were found.
\end{enumerate}
Once an object was located, it was not removed from the table. The time taken to find each object and the distance to the object were recorded. 
 
\subsubsection{Results}
\label{sec:FamResults}
\begin{figure}[htb] 
   \centering
   \includegraphics[width=0.65\textwidth]{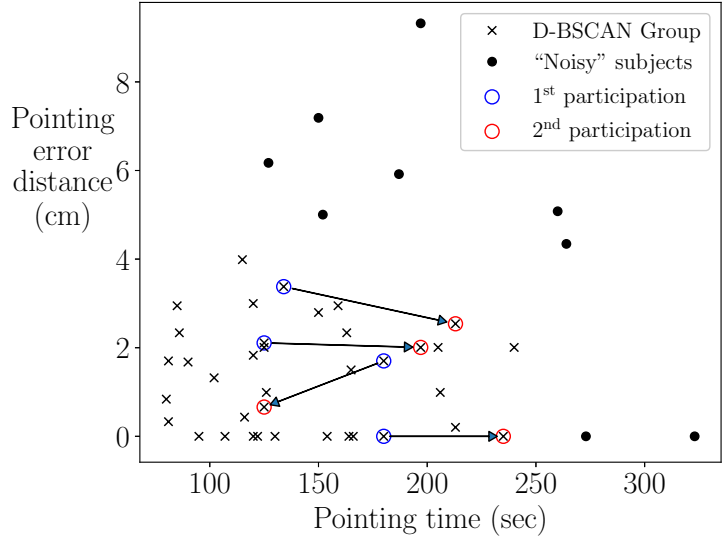}
   \caption{ 
   Familiarisation phase: Distribution of the mean distance between the index of the participants and the 3 objects vs the time taken by the participants to find the 3 objects. Black crosses are participants belonging to the same ``coherent'' group and black dots represent ``noisy'' participants found by 
    D-BSCAN. The four participants who participated twice are circled in blue for their 1\ts{st} participation and red for their 2\ts{nd} participation.}
   
   \label{fig:histandtimevsdist}  
\end{figure}

After the exploration phase, participants were able to locate the 3 objects with 
good accuracy. 
  Figure~\ref{fig:histandtimevsdist} gives the distribution of the participants according to the time they took to find the 3 objects and the mean distance between their index and the objects. The four participants who participated twice did not consistently perform better during their second participation: compared to their first participation, two were equally accurate and slower, one was slightly more accurate and slower, and one was more accurate and faster. This is likely due to the one-year time gap between their first and second participation. 
   Results are then compiled for the 42 participants who participated for the first time. 
  The mean required time to find the 3 objects over all 
 participants was $157\pm57$~seconds and the mean distance between the index and objects was $2.0\pm2.1$~cm.
 Based on the speed-accuracy tradeoff consideration~\cite{heitz2014speed}, the hypothesis that participants could have adopted two different approaches to complete the task depending on their willingness was tested: either participant is quick but not accurate or accurate but slower.
As is seen in Figure~\ref{fig:histandtimevsdist}, the hypothesis was not clearly valid. 
 The Density-Based Spatial Clustering of Applications with Noise (D-BSCAN)~\cite{DBSCAN1996} algorithm
 tries to 
 cluster 
points that have more than $minNeighbours$ surrounding neighbours 
within a distance $\epsilon$. 
 A D-BSCAN of the familiarisation data with the $\epsilon$ parameter set to $0.8$ and $minNeighbours = 5$\footnote{$minNeighbours = 5$ is the minimum number of neighbour points for a vector to be eligible for being in a cluster. (Fig.~\ref{fig:histandtimevsdist}). $\epsilon$ is defined with respect to the standardized data, i.e. data minus the mean and scaled to unit variance.}  
 confirmed that there is no clear separation between participants. 
 D-BSCAN found one general cluster of 37 quick and accurate participants 
 and 9 
 ``noisy'' participants who were either accurate but slow, quick but not accurate, or slow and not accurate.

\subsection{Navigation}
\subsubsection{Protocol}
Participants were asked to walk from one end of 
 the corridor (15 meters long and 6 meters wide) to the other without colliding with the objects from the scene (Fig.~\ref{fig:experiences} b,c). 
One pair each of chairs, garbage bins, small bags, and cardboard boxes were randomly placed in the corridor. Each participant had to make 
5 passes at their rhythm. We did not gave any recommendation. Participants were allowed to explore slowly or to go straight and fast.

 As participants were not blind, they could initially see the corridor/room and estimate the length of the run. We suspected this would impact the results, but observed that participants quickly forgot that estimation as they needed to concentrate on the use of the device while they navigated. We proceeded as follows to measure the participants' performances: 

\begin{enumerate}[label=(\roman*)]
\item Before each trial, we blindfolded the participants
and randomly moved the obstacles. 
\item We started the chronometer once they were ready.
\item One assistant remained close to the participant to take notes
(Fig.~\ref{fig:strategies}) about the strategy 
of the participant  and to make sure that they do not collide with objects. When participants detected an object on their way they needed to stop walking and scanning and have to tell the assistant what they perceived. The object was then reported by the assistant as seen. 
Walls or objects were classified as missed if the assistant had to intervene.
\item After the completion of a run, the chronometer was stopped and the time was noted.
\item Then, participants could rest or begin another run - restarting the process from step (i).
\item Once the 5 runs 
were completed, participants had to fill out a form to answer a list of questions related to the ease of use and to their confidence in the device (see Section~\ref{sec:questionnaire} for details).
\end{enumerate}

\subsubsection{Results}
\label{sec:NavResults}
 As in Section~\ref{sec:FamResults}, the four participants who participated twice did not consistently perform better during their second participation. 
Compared to their first participation and on average over their course completion time, two were faster, and two were slower. The average number of objects that they did not detect did not significantly change between their first and second participation. Again, this is likely due to the one-year time gap between their first and second participation. 
Among the 42 participants who participated for the first time, 
 2 did not complete the 5 courses and one did not understand the instructions. All other participants, except one, completed the courses with a mean time (over all courses) between $75$ and $362$ seconds. 
A last participant completed the courses with a mean time (over all courses) of $501$ seconds. This participant was identified as an outlier.
The number of missed objects and the time taken to complete the course were compiled on the remaining set of 38 participants who participated for the first time.

\begin{figure}[!htb] 
   \centering
   \includegraphics[width=0.65\textwidth]{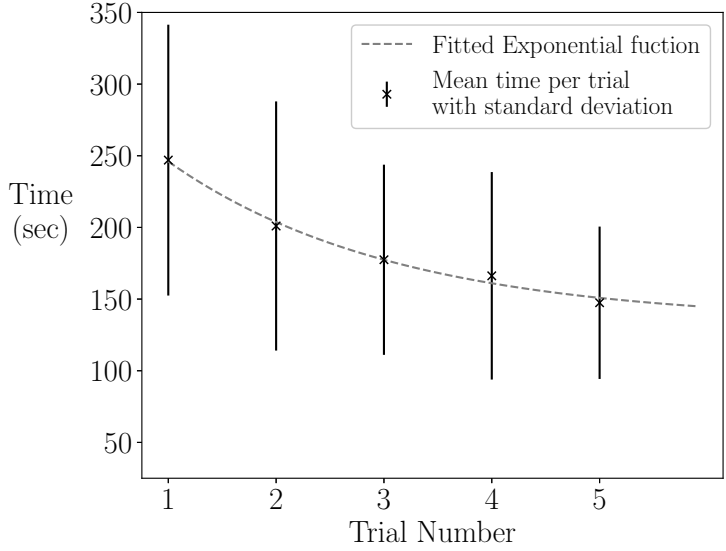} 
   \caption{Exponential decay function fits to the mean times per trial data
 }
   \label{fig:timeVsCourse1}
\end{figure}

 
 The global time performance of the participants measured with the mean times over all participants for each trial (Fig.~\ref{fig:timeVsCourse1}) significantly decreases and appears to follow an exponential decay law. 
 We used a non-linear least square optimization~\cite{kelley1999iterative} technique to fit a model of the time distribution as a function of the number of trials (Fig. \ref{fig:timeVsCourse1}).

We found that the evolution of the course mean times $\overline{Times}(n)$ over all participants follows an exponential decay law in relation to the number of trials $n$:

\begin{equation}
\overline{Times}(n) = 179.8 e^{-\frac{n}{2}} + 134.0
\end{equation}



  \begin{figure*}[!htb] 
   \centering
   \makebox[\textwidth][c]{\includegraphics[width=1.2\textwidth]{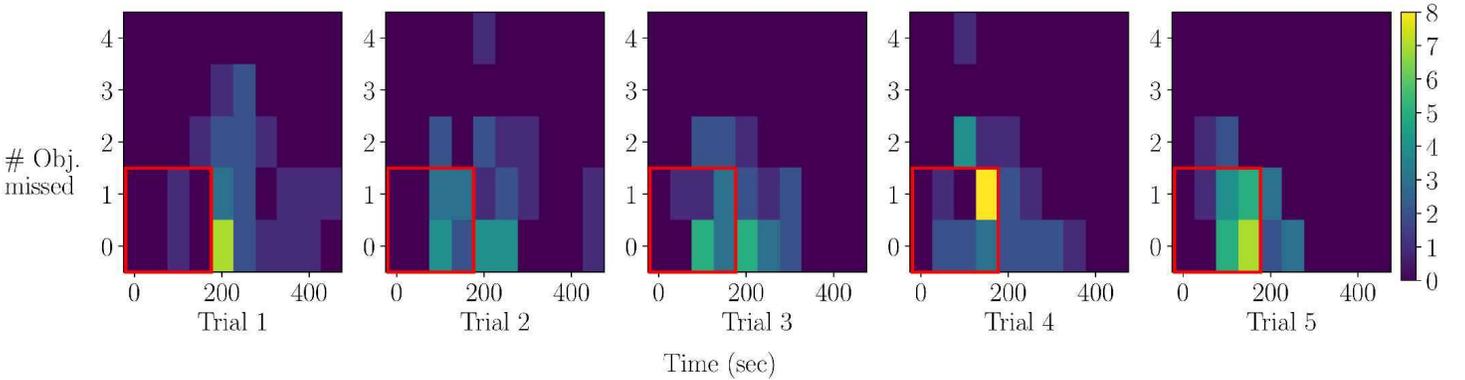} }
   \caption{2D histogram 
    for each trial of the distribution of the 42 participants with the number of objects missed versus the course completion time. Red outlines indicate participants who completed the course in less than 175~seconds and missed 0 or 1 object. A shift of the histograms to the bottom left is clearly observed.} 
   \label{fig:missedVsTimeHists}
\end{figure*}

Most participants significantly 
 improved their time across trials. 
More than 63\% of 
participants had 
their shortest time during the 4\ts{th} or the 5\ts{th} trial.
The average decrease time between the 1\ts{st} and best of the 4\ts{th} 
 or 5\ts{th} trial was of $107\pm78$~seconds.
We also defined the relative percentage improvement ${\%}_{improv}$ for each participant to measure the relative time improvement of the best of the two last trials relatively to 1\ts{st} trial as:
\begin{equation}
{\%}_{improv}= 100 - \frac{\min({t}_5,{t}_4)}{{t}_1} \times 100
\end{equation}
where $t_i$ is the time taken by participants to complete the $i\ts{th}$ trial. The mean of the relative percentage improvement (${\%}_{improv}$) over all participants was $40$\%.

\begin{figure}[htb] 
   \centering
   \includegraphics[width=0.65\textwidth]{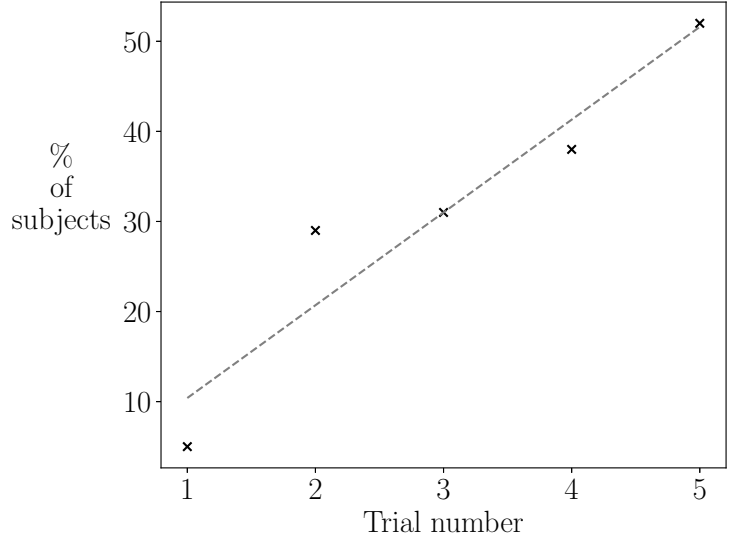} 
   \caption{Evolution of the percentage of participants who 
   missed 0 or 1 object and completed the course in less than 175~seconds. The dashed line shows a linear regression of the data.
   }
   \label{fig:linearRegRedGroup}
\end{figure}

Very few objects were missed during trials with a mean of $0.78$ a Standard Deviation (STD) of $0.87$ 
for all courses and participants. 
No statistically significant decrease in the number of missed objects 
across 
 trials was observed. Participants managed to efficiently use the system to detect obstacles 
 and became 
 more comfortable 
 and faster. Figure~\ref{fig:missedVsTimeHists} shows the distribution of the participants based on their course completion time and the number of objects they missed 
 per trial.
Participants who 
 completed the course in less than 175~seconds and who missed 0 or 1 objects are represented with a red outline. This group 
 comprises 5\% of the participants for the 1\ts{st} trial, 29\%, 31\% and, 38\%  respectively for the 2\ts{nd}, 3\ts{rd} and 4\ts{th} trial and 52\% for the 5\ts{th} trial (Fig.~\ref{fig:linearRegRedGroup}). 

\section{Questionnaire}
\label{sec:questionnaire}
\subsection{Description}
After the completion of the navigation task, 
each participant had to fill out a questionnaire comprising 11 questions displayed in table~\ref{tab:questionsDeLouis}. Three different types of questions were used: ``rating questions'', ``Yes or No questions'' and ``open questions''. For the ``rating questions'', participants had to provide a numerical answer comprised between 1 and 5, meaning respectively ``very easy'' and ``very difficult'' for Q1, 2, 4, 5 and, meaning respectively ``not afraid at all''' and ``very afraid'' for Q6. For the ``open questions'', participants were free to write whatever they wanted. The questionnaire was divided into three sections: one for the familiarisation, one for the navigation phase and one to provide general comments. 
 The questionnaire was 
 designed to quantify 
 participants's feelings and 
 get their feedback. 

\begin{table*}[htbp]
   \centering
      \caption{List of questions that were answered after the familiarisation and navigation tasks.}

   \begin{adjustbox}{center}
   \resizebox{1.\columnwidth}{!}{
   \begin{tabular}{@{} llr @{}} 
      \toprule 
      Question & Type \\
      \midrule
      Q1 -  How did you find the understanding of the device's manipulation? & Rating \\
      Q2 -  How did you find the task of finding the objects and to stay centered on them? & Rating \\
      Q3 -  What was your strategy to find the objects on the table ? & Open \\
      Q4 -  How did you find the task of pointing the object with your index finger ? & Rating \\
      Q5 -  How did you find the task of moving with the system ? & Rating \\
      Q6 - Were you afraid to hit the obstacles ? & Rating \\
      Q7 -  What was your strategy to move in the corridor ? & Open \\
      Q8 -  Did you forgot the iPhone during the experiment or were you aware that you were holding it ? & Yes or No \\
      Q9 -  Have you been bothered by the sound of the system ? & Yes or No \\
     Q10 - Would you be able to use the system during a long time (all day) ? & Yes or No \\
     Q11 - Have you any general comments on the system ? & Open \\
      \bottomrule 
   \end{tabular}
   }
   \end{adjustbox}
   \label{tab:questionsDeLouis}
\end{table*}

\subsection{Questionnaire Results}
 \label{sec:questResult}
Questionnaire results for the 46 participants
\footnote{The four participants who participated 2 times to the experiments answered the questionnaire twice.} 
are 
in table~\ref{tab:ratequestion} (``rating questions'') and in table \ref{tab:yesnoquesiton} (``Yes or No questions''). 
The operation of the SSD was in general easy to understand for the majority of the participants: 85\% of them answered 1 or 2 for the ease of understanding the manipulation of the device (Q1).
 The task of finding and pointing objects, which demanded high accuracy, was not very easy for the participants: 85\% and 91\% of the participants answered 2 or 3 respectively for Q2 and Q4 which corresponds to the ease of use for the familiarisation task. The navigation task was easier for most participants as 70\% answered 1 or 2 to Q5 which refers to the ease of moving with the system. 
They could approximately locate obstacles and move around them with less accuracy than the pointing task. 
 72\% of the participants answered 1 or 2 to Q6 which refers to the fear of use during navigation. 
 
Most participants did not find the sound annoying (63\% answered no 
 to Q9) but the majority (67\%) could not use the system for a long time: 
sighted participants had to do the task without vision which can be exhausting. Approximately half of the participants (54\%) 
did not realize they were holding the system (Q8) during the navigation task. 

Answers to Q3 and Q7 on the strategies used by the participants combined with the author's observations revealed various strategies for the familiarisation and navigation phases. 
During the familiarisation phase, two strategies were 
observed.
 One can be 
 named as \emph{all or nothing} in which participants scan the space very close to the table and listen to sounds without paying attention to the spatialization of the sound. When an object is detected very close to the system, many cells from the grid (Fig.~\ref{fig:positionnementDeLaGrille}) are activated and multiple sounds are triggered simultaneously. Participants then know that the object is right under the camera device.
The other strategy could be named \emph{coarse to fine}: participants scan the entire table for an overview. With this strategy, when an object is detected far from the system, only one cell is activated, and one sound is triggered, which makes the interpretation of the sound position easy. Participants then try to locate each object by getting closer to the table, guided by the sound spatialization. 

\begin{figure}[htb] 
   \centering
\begin{subfigure}{0.99\linewidth}
    \centering \includegraphics[width=0.5\linewidth]{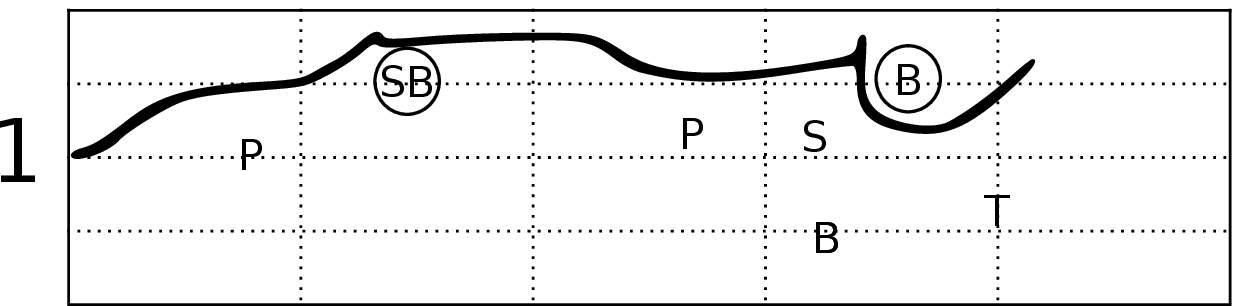}
    \centerline{(a)}
 \end{subfigure}
 \begin{subfigure}{0.99\linewidth}
 \centering \includegraphics[width=0.5\linewidth]{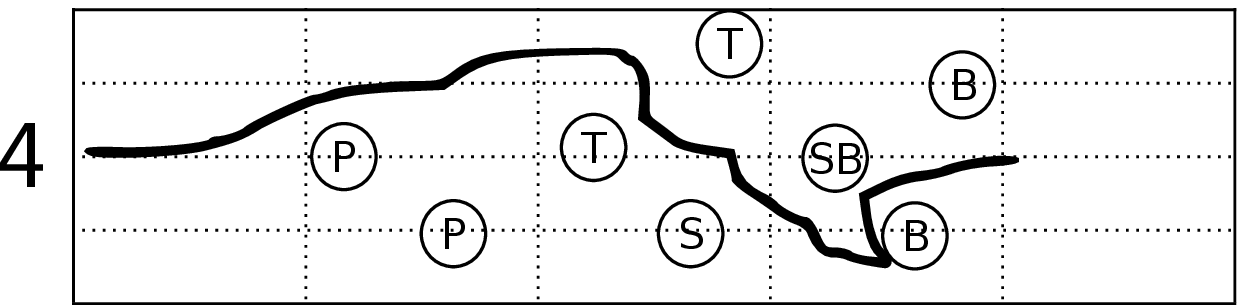}
 \centerline{(b)}
 \end{subfigure}
 
   \caption{Examples of strategies: 2 different strategies for the same participant. P is a garbage bin, 
S a bag, SB a blue bag, B a cardboard box and T a chair. Circled letters indicate found objects. (a) Going straight through the corridor and finding only objects that are on the path, (b) finding as many objects as possible while going through the corridor.}
   \label{fig:strategies}
\end{figure}

During the navigation phase, we observed 3 strategies.
The first could be named \emph{follow the silence} (Fig.~\ref{fig:strategies} a) in which participants proceed step by step (stop walking, scan the environment to find a silent direction, and then move 1-2 meters ahead in the direction of the silent area). 
The second strategy was to alternate between stop and scan actions to walk from obstacle to obstacle (Fig.~\ref{fig:strategies} b). This strategy could be named \emph{follow obstacles}. 
The last strategy, that could be named \emph{scan while walking}, was to continuously scan the space while moving and stop only if there is an obstacle in close proximity.


\begin{table}[!htbp]
     \centering
        \captionof{table}{Rate questions}

 \begin{tabular}{cccccc}
    \toprule
                 & Q1  & Q2 & Q4 & Q5 & Q6 \\ \midrule
    Mean & 1.9   & 2.5   & 2.6 & 2.1  & 1.9 \\
    STD & 1.1   & 0.8   & 0.9 & 1.6  & 1.1 \\ \bottomrule
  \end{tabular}
    \label{tab:ratequestion}
   \end{table}	
   
   \begin{table}[!htbp]
      \centering
         \captionof{table}{Yes / No questions}

 \begin{tabular}{cccccc}
    \toprule
                 & Q8  & Q9 & Q10  \\ \midrule
    Percent Yes & 54 \%   & 37 \%  & 33 \% \\
    Percent No & 46 \%   & 63 \% & 67 \% \\ \bottomrule
  \end{tabular}
       \label{tab:yesnoquesiton}
   \end{table}

\section{Discussion}
\subsection{Comparison with similar works}

\begin{table*}[htpb]
\centering
\caption{Comparison of localisation error distances between SSDs}
\begin{adjustbox}{center}
\resizebox{1.\columnwidth}{!}{
\begin{tabular}{@{}ccccccc@{}}
\toprule
SSD                 & \begin{tabular}[c]{@{}c@{}}The vOICe\\ \cite{auvray2007learning}\end{tabular}  & \begin{tabular}[c]{@{}c@{}} PSVA,\\ \cite{Renier2010}\end{tabular} &\begin{tabular}[c]{@{}c@{}} PSVA,\\ \cite{Renier2010}\end{tabular} &\begin{tabular}[c]{@{}c@{}}The Vibe\\ \cite{hanneton:hal-00577751}\end{tabular} & \begin{tabular}[c]{@{}c@{}}The Vibe\\ \cite{hanneton:hal-00577751} \end{tabular} & \textbf{See Differently} \\ \midrule
Localisation error (cm) &    $7.8\pm5.1$    &      $19.6$  &    $12.3$  &  $6.48\pm5.00$                              &                   $6.00\pm3.1$                                  &                                        $\boldsymbol{2.1\pm2.0}$    

\\ \midrule
Training time &  5 hours   &     17 hours  &     17 hours   &      \begin{tabular}[c]{@{}c@{}}Not\\ reported\end{tabular}   &           \begin{tabular}[c]{@{}c@{}}Not\\ reported\end{tabular}                &                                        \textbf{10 minutes}
\\ \midrule
Constraints &     \begin{tabular}[c]{@{}c@{}}Stationary elbow\\ Hand-held\end{tabular}    &     \begin{tabular}[c]{@{}c@{}}Free movement\\ Head-mounted\end{tabular}  &     \begin{tabular}[c]{@{}c@{}}Free movement\\ Head-mounted\end{tabular}  &       \begin{tabular}[c]{@{}c@{}}Stationary elbow\\ Left hand-held\end{tabular}     &         \begin{tabular}[c]{@{}c@{}}Stationary elbow\\ Right hand-held\end{tabular}      &               \begin{tabular}[c]{@{}c@{}}\textbf{Free movement}\\ \textbf{Hand-held}\end{tabular}    
\\ \midrule
Participants &   \begin{tabular}[c]{@{}c@{}}Blindfolded\\ sighted\end{tabular}   &       \begin{tabular}[c]{@{}c@{}}Early\\ blind\end{tabular}    &       \begin{tabular}[c]{@{}c@{}}Blindfolded\\ sighted\end{tabular}    &      \begin{tabular}[c]{@{}c@{}}Blindfolded\\ sighted\end{tabular}   &           \begin{tabular}[c]{@{}c@{}}Blindfolded\\ sighted\end{tabular}                &                                         \begin{tabular}[c]{@{}c@{}}\textbf{Blindfolded}\\ \textbf{sighted}\end{tabular}  
\\ \bottomrule
\end{tabular}
}
\end{adjustbox}
\label{table:poitingerror}
\end{table*}

We evaluated the learning curve of 
46 blindfolded sighted participants using the \emph{See Differently} SSD for objects localisation and navigation.
The lack of a standardized protocol to evaluate SSDs and the various purposes of publications studying SSDs make an objective comparison challenging. To gain an insight into the potential of the SSDs to provide the ``where information'', we reported here experiments conducted with protocols that involved an object location task or a navigation task in the 3D ``real world''. We only reported experiments conducted with a visual to auditory SSD that do not use other sensors than a 2D camera, and where a measure of the localisation error or the time taken to complete the navigation task are provided. It would be interesting in future work to compare the SSDs with strictly identical protocols.
For the location task, participants had to find 3 objects on a table. They were able to do the task after only 10 minutes of training 
and seemed generally better at pointing with \emph{See Differently}~(as seen inTable~\ref{table:poitingerror}.

For the navigation task with \emph{See Differently}, participants had to navigate through a corridor while avoiding obstacles randomly placed on the ground. They improved their averaged time by 104 seconds over 5 trials 
while avoiding almost every obstacles (averaged number of missed obstacles of $0.78$ with a STD of $0.87$). They needed far less training (approximately 10 minutes) than in similar experiments conducted 
with \emph{the vOICe} (approximately 3 hours of training)~\cite{Brown2011} and with~\emph{The Vibe} (training over 4 days) ~\cite{durette:inria-00325414}.
With~\emph{The Vibe}~\cite{durette:inria-00325414}, participants had to complete 
a U-shaped track (60 meters) in a car park several times over 4 days 
 (3 courses per day). 
They did not have to avoid obstacles and were simply told when they 
were out of the track. 
As a comparison indicator for the two experiments, we computed how often the researchers had to intervene to stop participants (referred to as missed objects in our experiments). It was on average every 9 meters~\cite{durette:inria-00325414} with~\emph{The Vibe} and every 19 meters in our experiments with~\emph{See Differently}.
~\cite{Brown2011} also conducted navigation experiments with obstacles with~\emph{the vOIce} but did not provide the number of collisions (or the number of times the researchers had to intervene). 

The fast learning 
 and good performance observed for the 2 tasks with the very simple sonification algorithm of \emph{See Differently} suggests that the training using location task to the use of SSDs can be quick. This is done in our work by not overloading the user's audition and providing only simple location features (this system is not doing any recognition). 
The motivation to avoid overloading the user's audition is supported by
~\cite{Brown2011} who conducted experiments on image recognition with \emph{the vOICe} to address the problem of the sensory capacity limitation in sensory substitution system. They observed that drastically reducing the amount of the sonified  information at the input of \emph{the vOICe} system (from 128x128 pixels to 8x8 pixels images) didn't affect the performance of an image recognition task. They later showed that participants better recognize objects with \emph{the vOICe}~\cite{brown2016audio}  when 
 images were divided into
  two sub-images before being converted into sound. 

Another important aspect of \emph{See Differently} is that users interpret the soundscapes and elaborate their own strategy in the way they use the device. This versatility is highly important to reduce the learning time and accommodate the device to any user's strategy. Users were also able to elaborate their own strategy with \emph{the vOICe}~\cite{auvray2007learning}, the \emph{PSVA}~\cite{Renier2010} or \emph{The Vibe}~\cite{hanneton:hal-00577751,durette:inria-00325414}, though over a much longer training period. 

There is a trade-off between the abstraction level of the provided visual information and the task for which the device would be used (for location: ``where'' and/or for identification: ``what''). Too much abstraction could lead to a lack of generalization and interpretation by the user. 
On the other hand, encoding all raw image pixels without any interpretation by the device leads to extensive training times. The proposed approach is a tradeoff for which the system provides the ``where information'' in a simple and efficient way, thus reducing the training time.

\subsection{Location task with a congenitally blind participant}
Although this work has not been designed for early blind people, it is interesting to know the limits of~\emph{See Differently} for these people. We therefore conducted a preliminary experiment with a congenitally blind participant. 
The experiment comprised the pointing task with the same protocol as described in Section~\ref{sec:FamProtocol} repeated 4 times, and the answer to the questionnaire described in table~\ref{tab:questionsDeLouis}. 
The task was also approved by the same ethical committee and certificate under reference number 2014-85/Rouat. Compared to the blindfolded sighted participants, the 
subject did not immediately understand that the sound was spatialized with reference to the screen of the device. After 
10 minutes of familiarisation, the participant understood the sonification strategy well and successfully located the 3 objects on the table. The mean required time to find the 3 objects over the 4 trials was $168\pm13$ seconds and the mean distance between the index and the objects was $2.5\pm0.9$ cm. Therefore, the subject belongs to the cluster of participants considered to be quick and accurate in Section~\ref{sec:FamResults}. It is interesting to note that, across the 4 trials, the participant first used the \emph{all or nothing} strategy 
 and then employed the \emph{coarse to fine} strategy 
 which the subject found to be more efficient. These preliminary results show significant potential for the use of the system by early blind people (at least for the location of objects).

\section{Conclusion}



SSDs are very promising for use by blind or low vision people. However, the extensive training time required or the complex and expensive technologies needed are a barrier to their widespread use. 
Our contribution is a proof of concept to show
that simple and effective SSDs can be designed with fast training (at least for the ``where" modality). 
 
 Future work will focus on improving the image processing algorithm, adding new functionalities and conducting other experiments with both blind and sighted people.
For the image processing algorithm, the threshold used in equation~\ref{eq:OE_AlgoThresh} should be adapted according to the brightness. As the goal here is not to propose a new image processing algorithm, we conducted our experiments in an indoor environment where the brightness has a small dynamic range. We plan for the future to improve 
and compare our image processing algorithm to other methods of salient feature detection. 
Further experiments will include a study of the reliability of the system with respect to the distance and an in-depth comparison of our sensory substitution strategy with other strategies.
 
Future work will also focus on adding a second functioning mode (``what'' mode) with the same will of keeping a short training period. For the ``what'' mode we will extract higher level features like contours, shapes and some form of categorization with unsupervised machine learning techniques that do not require too much data to train on. It will allow the user to identify visual features like shapes, sizes and eventually colors of salient parts of the visual scene. We again would like the user's brain to interpret the scene and not the system. As we have shown in this work, systems that leverage user perception to easily interpret the scene show significant potential for fast learning and flexibility to adapt to different user strategies.


\appendix
\section{Neural network}
\label{appendix:neuralNet}
We provide here the details for the neural network as introduced in the work of~\cite{lescalISSPA2012}. 

 \subsubsection{The Network}
 Synapses connect each neuron to its 8 neighbours (including diagonal connections). 
Connecting weights $w_{ij}$ between neurons $i$ and $j$ are bidirectional and given by:
\begin{equation}\label{eq:OE_SynapseWeight}
w_{ij} = w_{ji} = f\left(|p_{i}-p_{j}|\right)
\end{equation}
where $p_{i}$ is the pixel greyscaled value of neuron $i$. $|\; \;|$ is the absolute value and $f\left(\right)$ is given by: 
\begin{equation}\label{eq:OE_PixelDiffFunct}
f\left(x\right) =1 - \frac{x}{MAX}
\end{equation}
with $MAX=255$. 
The larger is the difference between pixels, the smaller are the weights. Weights are comprised between $0$ and $1$.

\subsubsection{Neurons}
Each neuron has a feedback connection 
and remembers its previous state. An iteration of the algorithm passes through the list of all neurons in the network and updates their state $s_{i}[n]$. 
\begin{enumerate}[label=(\roman*)]
\item The update is done by cumulating the weighted states of the neighbouring neurons.

For instant $n$:
\begin{equation}\label{eq:OE_AlgoUpdate}
\hat{s}_{i}[n] = \frac{s_{i}[n-1] + \sum\limits_{j}{w_{ij} . s_{j}[n-1] 
}}{9},
\end{equation}
where $s_{i}[n-1]$ is neuron $i$'s state at instant $n-1$ of the algorithm and $\hat{s}_{i}[n]$ is the estimated state before thresholding and the divisor is set to 9 to normalize the state of the neuron between $0$ and $1$. Because of the recursive component in equation \ref{eq:OE_AlgoUpdate}, gradients, contours and textures diffuse and propagate inside the network to expand the mask. 
 It was observed in~\cite{lescalISSPA2012} that a maximum of 4 iterations is sufficient for most images.
 
\item The activation function of the neurons is the identity transform when the estimated state $\hat{s}_{i}[n]$ is sufficiently large. Otherwise, the state is forced to 0 (equation~\ref{eq:OE_AlgoThresh}).

The new state $\hat{s}_i[n]$ from equation~\ref{eq:OE_AlgoUpdate} is compared with a threshold (equation~\ref{eq:OE_AlgoThresh}) and set to 0 if it is smaller than an empirically-determined threshold.
\begin{equation}\label{eq:OE_AlgoThresh}
s_{i}[k] \leftarrow
\left\{
\begin{array}{ll}
0 & \mbox{if } s_{i}[k] \leq THRESH \\
\hat{s}_{i}[k] & \mbox{otherwise}
\end{array}
\right.
\end{equation}
Neurons with the smallest state are in highly textured and contrastive areas.

\item
If a maximum number of iterations -- fixed by the user -- is reached the algorithm exits, otherwise, $n \leftarrow n+1$ and then go to step (i) to estimate the next value of the state of each neuron.

\item In the end, the neuron's states (comprised between 0 and 1) are used to mask and enhance the areas of interest in the image.
\end{enumerate}

For real time processing on a slow device, 
 only one iteration can be used ($n=1$) with initial states $s_i[0]$ set to 1. $THRESH$ is equal to $0.112$. 
After running the network, pixels from the  inactive neurons (i.e. $s_{i}[n] = 0$) are identified as points being parts of areas of interest (salient points), i.e. the areas to be encoded into sounds.
In this work, only one iteration was necessary.
 Equations~\ref{eq:OE_SynapseWeight} and~\ref{eq:OE_AlgoUpdate} become: 
 \begin{equation}
 \hat{s}_{i}[n] = \frac{(1+ \sum_{j} f(|p_{i} - p_{j}|))}{9}
 \end{equation}
 

\section*{Author Contributions}
L.C., S.W., and J.R. conducted the experiments, analyzed the results and wrote the paper; J.R. and S.W designed and implemented the 3 mapping modes of \emph{See Differently}; J.R. designed and implemented the neural network on a preliminary implementation of \emph{See Differently}.

\section*{DeclarationofCompetingInterest}
None.

\section*{Acknowledgments}
D. Lescal for a preliminary implementation and design of \emph{See Differently}. 
The participants, L. Bachatène, V. Bharmauria. S. Cattan, N. Chanauria, S. Molotchnikoff, E. Plourde and the members of the NECOTIS research group for testing and providing feedback. Franco Lepore and Patrice Voss for stimulating and fruitful discussions for the planning of the navigation protocol.
 This work was supported by FRQNT team grants program and the NSERC-CRSNG discovery grants program.

\bibliographystyle{unsrt}  
\bibliography{references}

\end{document}